\documentclass[12pt,preprint]{aastex}

\newcommand\beq{\begin{equation}} 
\newcommand\eeq{\end{equation}}

\def\erg{$\rm erg\,s^{-1}$} 
\def\fxu{{$\rm erg\ cm^{-2}\ s^{-1}$}}

\newcommand{\gta}{\gtrsim}

\def\lesssim{\,
\lower2truept\hbox{${<\atop\hbox{\raise4truept\hbox{$\sim$}}}$}\,}
\def\gtrsim{\,
\lower2truept\hbox{${> \atop\hbox{\raise4truept\hbox{$\sim$}}}$}\,}

\received{} 
\accepted{} 

\shorttitle{ULXs in NGC~4565}
\shortauthors{Wu et al.} 

\begin{document}

\title{ The nature of ultra-luminous X-ray sources in NGC~4565}

\author{ 
H. Wu\altaffilmark{1}, 
S.J. Xue\altaffilmark{1},
X.Y. Xia\altaffilmark{1,2},
Z.G. Deng\altaffilmark{1,3}
and S. Mao\altaffilmark{4}
}

\altaffiltext{1}{National Astronomical Observatories, 
    Chinese Academy of Sciences, Beijing 100012, China }

\altaffiltext{2}{Department of Physics, Tianjin Normal University, Tianjin 300074, China}

\altaffiltext{3}{Graduate School, Chinese Academy of Sciences, 
    Beijing 100080, China}

\altaffiltext{4}{Univ. of Manchester, Jodrell Bank Observatory,
          Macclesfield, Cheshire SK11 9DL, UK}

\authoremail{wu@vega.bac.pku.edu.cn}

\begin{abstract}
We report the optical identifications of two X-ray luminous point sources
in the spiral galaxy
NGC~4565 based on archive data of Chandra and the Hubble Space
Telescope. The central X-ray point source,
RXJ1236.3+2559, is found to be the nucleus of NGC~4565 with an
X-ray luminosity of $L_{x}\approx 
4.3\times10^{39}$ \erg. We show that its multi-waveband properties are
consistent with it being a low-luminosity active galactic nucleus.
A faint optical counterpart with  $B \approx 25.1$ and
$I \approx 24.0$ was identified for the off-nucleus X-ray point source,
RXJ1236.2+2558.  Its extinction-corrected B magnitude is estimated to
be 24.5. The X-ray to optical flux ratio  ($f_X/f_{\rm B}$)
is about 540. From the optical and X-ray 
properties we argue that RXJ1236.2+2558 is an ultra-luminous X-ray compact
source with $L_{x}\approx 6.5\times10^{39}$\erg. The source is probably located in 
a faint globular cluster at the outer edge of NGC~4565's bulge.

\end{abstract}


\section{Introduction}

Recent observations present new evidence for the existence of off-nuclear
X-ray luminous point sources in nearby spiral galaxies, starburst galaxies and 
elliptical galaxies. X-ray luminosities of these sources 
are of $10^{39}-10^{40}$ \erg\, in the 0.5-10 keV band, which are significantly 
larger than the Eddington luminosity of a $1 M_{\odot}$ object and are 
intermediate between classic X-ray binaries (with $L_{x}\approx 10^{36}-10^{38}$ 
\erg) and active galactic nuclei (with $L_{x}\gta 10^{41}$ \erg). These X-ray 
luminous, non-AGN, non-supernova remnant compact objects have been called 
ultra-luminous compact X-ray sources (ULXs) (Fabbiano 1989; Matsumoto et al. 
2001; Makishima et al. 2000; Roberts et al. 2002; Sarazin, Irwin \& 
Bregman 2001; and references therein). The nature of ULXs is far from
clear. Current models include intermediate-mass ($\geq 50 M_{\odot}$) 
black holes (Madau \& Rees 2001) and
significantly beamed binary systems (e.g., K\"ording et al. 2002;
King et al. 2001 and references therein; for more, see the discussion).
 
To understand the nature and origin of ULXs, it is important to explore their 
multi-waveband properties and their environments. Given the superior 
spatial resolution of Chandra, it is possible to identify the 
optical counterparts of ULXs in nearby galaxies. Such studies are underway. The 
ULX in the nearby starburst galaxy M82 (M82 J095550.2+694047) was
proposed to be an intermediate mass black hole (BH) (100-1000 $M_{\odot}$) in an intense starburst
region, from Chandra and millimeter observations (Matsumoto et al. 2001; Matsushita et al. 
2001). Recently, Roberts et al. (2001, 2002) found that two ULXs in NGC 5204 
are located in cavities of emission-line gas nebulae while six ULXs in
the interacting pair galaxies NGC4485/4490 are located in star forming regions. Wang (2002) also 
found that the ULX X-9 in M81 is related to a shell-like optical nebula in the tidal 
arm of M81. In another study, Liu, Bregman \& Seitzer (2001) found several optical counterparts 
of ULXs in M81, M51 and NGC2403; these optical counterparts
are all quite blue with properties consistent with O stars. Most recently, 
Pakull \& Mirioni (2002) reported their optical survey results for ULXs in 11 
nearby galaxies. At the position of several ULXs, they found emission line 
nebulae of a few hundred parsecs in diameter, which often show both low and 
high ionization emission lines. On the other hand, ULXs have also been found
in globular clusters of giant elliptical galaxy NGC 1399 (Angelini, Loewenstein \& Mushotzky 2001).
It is not yet clear whether ULXs located in globular clusters
have different properties from those in star formation regions.

NGC 4565 is a nearby edge-on spiral galaxy of type Sb. It is at
a distance of 10 Mpc based on surface brightness fluctuations and 
planetary nebulae (Jacoby, Ciardullo, \& Harris 1996 and references
therein).
Its surface brightness profile has been extensively studied (Wu et al. 
2002 and references therein). Using ROSAT and ASCA data,
Mizuno et al. (1999) found two bright point-like X-ray sources in the
field of NGC 4565. The fainter one is positionally coincident with 
the nucleus of NGC 4565 and would normally be considered
as a low-luminosity active galactic nucleus (AGN). However, Mizuno et al. argued
that it is a ULX instead because the neutral
hydrogen column density determined from X-ray spectral fitting is too
low for a central AGN. On the other hand, the brighter source is
off-center and about 2 kpc above the galaxy 
disk if it is associated with NGC~4565.
Spectral fittings of ASCA data show that a multi-color disk (MCD,
see e.g., Mitsuda et al. 1984) 
model provides the best fit for both sources and their 0.5-10 keV luminosities 
are higher than $10^{39}$ \erg. Therefore, Mizuno et al. (1999) concluded that 
both sources are ULXs. However, due to its low spatial resolution,
ASCA cannot separate these two X-ray bright sources well, so their
conclusion needs to be carefully verified.
 
In this paper, we present the optical identifications for these two X-ray point 
sources in NGC~4565 using the archive data from a Chandra snapshot survey 
and from the Hubble Space Telescope (HST) observations. The outline of the paper is as following. In 
\S 2, we briefly describe the observational data we used and the data reduction.
In \S 3, we discuss the optical identification of the two X-ray bright 
sources in NGC~4565, and their properties.
In \S 4, we summarize our results and further discuss the nature of these two
point sources.

\section{Observations and Data Reduction}

The X-ray data for NGC~4565 are retrieved from the Chandra archive\footnote{
The Chandra Data Archive (CDA) is part of the Chandra X-Ray Observatory 
Science Center (CXC) which is operated for NASA by the Smithsonian
Astrophysical Observatory}.
The data are from a Chandra snapshot survey 
for a volume-limited sample of 25 low-luminosity active galactic nuclei 
candidates\footnote{PI: Garmire, CHANDRA OBSID 404}. The survey was
taken using the ACIS-S instrument on board the Chandra X-ray observatory. The observations of 
NGC~4565 were performed on June 30, 2000 with a total exposure time of about 2~ks. 
The data reduction was carried out using the Chandra software package, CIAO2.1, 
with the latest calibration database, CALDB2.8. Both X-ray 
bright sources in NGC~4565 
are clearly detected by 
Chandra ACIS-S3. The photon counts are 121 and 269 for the central
(RXJ1236.3+2559) and the off-center (RXJ1236.2+2558) sources in a circular region of
radius 2.5$''$ and 3$''$, respectively. Even under the sub-arcsecond resolution of 
Chandra ACIS-S, these two X-ray sources in NGC~4565 remain unresolved.
 
HST WFPC2 images in two bands (F450W and F814W) were retrieved using
NED \footnote{
The NASA/IPAC Extragalactic Database (NED) 
is operated by the Jet Propulsion Laboratory, California Institute of
Technology, under contract with the National Aeronautics and 
Space Administration}. 
The total integration times of the F450W 
and F814W images are 600s and 480s, respectively. More detailed information can 
be found in Kissler-Patig et al. (1999). To augment the HST observations,
we also performed narrow band H$\alpha$ imaging for NGC~4565 using the 2.16m
telescope and the 60/90cm Schmidt telescopes at the Xinglong Station of NAOC on May 18,
2001 and  April 13, 2001, respectively. The Central source (RXJ1236.3+2559) was
clearly detected in H$\alpha$. All the optical data
reductions are performed using the IRAF package through standard procedures.

\section{Results}

\subsection{Optical Identifications of Two Bright X-ray Point Sources }

Fig. 1 shows the HST WFPC2 F814W image of the bulge of NGC~4565. An image of the whole galaxy
is shown in the bottom-right inset . The bottom-left and top-right insets show the optical 
counterparts of the two Chandra 
X-ray bright point sources. The pluses in the insets show the Chandra X-ray source
positions.  It is clear from Fig. 1 that
the optical nucleus and the off-center faint source are located within a
0.5\arcsec~ error circle of the two Chandra bright point sources. To confirm the 
existence of the faint optical counterpart of RXJ1236.2+2558, we carefully 
checked the six pre-combined HST images (three F450W and three F814W 
images) individually. The faint optical counterpart appears in all of the six 
images. Radio continuum observations at 6cm and 20cm by the Very Large Array (VLA) for NGC~4565
detected the nuclear source (Ho \& Ulvestad 2001). The positional
accuracy of the radio 
observations is about 0.1$''$. The cross in the bottom-left inset of Fig. 
1 shows the radio position of the nuclear source. Table 1 lists the positions of 
RXJ1236.3+2559 and RXJ1236.2+2558 determined by HST, Chandra and VLA. 

Notice that, in Table 1 the optical positions for both RXJ1236.3+2559 and 
RXJ1236.2+2558 are about 0.3$''$ north from the corresponding Chandra positions. 
Such a shift can arise due to the uncertainty in the absolute astrometry
in both HST and Chandra coordinate systems.  After correcting this shift, the 
HST and Chandra positions agree very
well with a relative positional error of about 0.1$''$.

\subsection{RXJ1236.3+2559: a low-luminosity active galactic nucleus in NGC~4565}

The high resolution observations for NGC~4565 by Chandra, HST and VLA
determine accurately the positions of the nuclear source 
to about 0.1$''$ (corresponding to a physical scale of about 5pc). These observations
allow us to investigate the properties of RXJ1236.3+2559 
and establish its physical nature. 

The background-subtracted X-ray spectrum of RXJ1236.3+2559 was extracted with an aperture
of a radius of 2.5$''$ where we have determined the background using
a local source-free annulus between radius 5$''$ and 15$''$.
At the present flux level, the 
pile-up fraction of the count rate for both 
X-ray sources is expected to be much less than 10\%,
so we ignored the pile-up effect in our spectrum extraction.
Not surprisingly, 
the small number of detected photons from the Chandra snapshot survey observation 
prevents us from discriminating different models for the X-ray spectra of 
RXJ1236.3+2559. Indeed, both an absorbed power-law model
and an absorbed multi-color disk blackbody (MCD) model provide an acceptable fit for
the spectrum. The fitting parameters (including the
$\chi^2$ per degree of freedom) for both models are listed in Table 2;
the top panel of Figure 2 shows the best absorbed power-law fit.

 For the absorbed power-law model, the photon index is 
$\Gamma=1.6$ with a neutral hydrogen 
column density of $N_{\rm H} = 1.1 \times$10$^{21}$ cm$^{-2}$, which is consistent with the result 
of Mizuno et al. (1999) who obtained
$N_{\rm H} \leq$2.0$\times$10$^{21}$ from ASCA observations. As
pointed out by Vogler, Pietsch, \& Kahabka (1996), this value is below the HI column density 
of about 5$\times$10$^{21}$ cm$^{-2}$ derived from an HI map 
of NGC~4565 at a resolution of 40\arcsec\, (Rupen 1991).
As NGC~4565 is an edge-on spiral galaxy, Mizuno 
et al. argued that the column density along the galaxy disk to the nucleus of 
NGC~4565 should be at least 1$\times$10$^{22}$ cm$^{-2}$. Based on this,
they rejected the possibility of RXJ1236.3+2559 being a
low-luminosity AGN. Instead, they suggested that it
is a ULX located at the near-side of this disk galaxy. 
 
However, this conclusion is uncertain because the interstellar medium
is known to be clumpy, so the HI column density toward a particular
line of sight can be lower or higher than the average value. So we carefully investigated the extinction 
in the nuclear region of NGC~4565 using the color determined from
the HST F814W and F450W images.
We find that the nucleus is located in a 
low reddening area just outside the dust lane. Therefore, it seems plausible 
that the HI column density in the nucleus is lower than the value 
estimated from the low resolution HI map. This leads us to postulate that the
X-ray emission of RXJ1236.3+2559 is from a low-luminosity AGN. 

To further prove the low-luminosity AGN hypothesis for RXJ1236.3+2559, we collected 
all the available multi-waveband information for this source from the literature. The 
optical photometry is obtained from Kissler-Patig et al. (1999). The
measured I and B band nuclear magnitudes with an aperture size of 0.5$''$ are 17.2
and 19.5, respectively. The near-infrared J, H and K bands photometry are from 
Rice et al. (1996). A 5$''$ aperture size were used for photometry in these 
three bands and the measurements were transformed into flux densities.  The
1.4GHz--15GHz radio data are from VLA observations by Nagar, Wilson \& Falcke (2001)
and Ho \& Ulvestad (2001) within an aperture size of about 0.5$''$. 
The ultraviolet data are
from the HST Faint Object Camera (Maoz et al 1996), which gives the
integrated 2300$\rm \AA$ flux with an aperture of $22''\times 22''$;
this can only be considered as an upper limit of the nucleus flux due to
the contribution from other stars in the bulge.

All these data are 
presented in Table 3 and the spectral energy distribution for the nucleus of 
NGC~4565 is shown in Fig. 3. As a comparison, we also plotted the spectral
energy distribution for the
low-luminosity active galactic nucleus in M81 (dotted line) and NGC~4579 (solid
line) from Ho (1999). As one can see, the spectral energy distribution of 
NGC~4565 from radio, optical to X-ray is quite similar to those of M81 and 
NGC~4579, though differences in the near infrared bands exist. 
As mentioned above, due to the larger aperture adopted for the photometry of J, H and K bands,
their values can only be regarded as upper limits. As suggested by Ho (1999), 
the contamination from starlight in the near-infrared greatly exceeds those in
other bands. The resemblance of the spectral energy distribution for the
nucleus of NGC~4565 with those of low-luminosity
AGNs is another piece of evidence that it is a low-luminosity active galactic nucleus.

Furthermore, we studied the relation of H$\alpha$ and X-ray emissions. 
The H$\alpha$ flux for RXJ1236.3+2559
is 2.4$\times10^{-14}$ erg s$^{-1}$ cm$^{-2}$ within an aperture of 
radius 5$''$ which corresponds to an H$\alpha$ luminosity of 2.8$\times$10$^{38}$ 
erg s$^{-1}$. As the X-ray luminosity of nuclear source of NGC~4565 is 4.3
$\times$10$^{39}$ erg s$^{-1}$, the ratio of the X-ray luminosity to
the H$\alpha$ luminosity for NGC~4565 is 15.3, which is just the median ratio
(15) for low-luminosity AGNs (Ho et al. 2001). Therefore, we conclude that
RXJ1236.3+2559 is a low-luminosity AGN. 

\subsection{RXJ1236.2+2558: a ULX in a Stellar Cluster}

\subsubsection{The Optical and X-ray Properties of RXJ1236.2+2558}
 
As shown in Fig. 1, the bright X-ray point source RXJ1236.2+2558 lies at the 
outskirts of the bulge of NGC~4565. Photometry measurement reveals that the optical 
counterpart of RXJ1236.2+2558 is quite faint with
$B=25.1 \pm 0.4$ and  $I=24.0 \pm 0.3$ within an aperture of 0.5$''$.

The X-ray spectrum was extracted with an aperture size of 3$''$ and with
the local source-free background in an annulus subtracted. 
Both an absorbed power-law model and an absorbed MCD model can fit the
spectrum. The bottom  panel of Fig. 2 shows the best MCD model spectrum of RXJ1236.2+2558.
All the fitting parameters are listed in Table 2. Based on the MCD model 
and assuming RXJ1236.2+2558 is in NGC~4565, the absorption-corrected soft X-ray (0.1-2.4 keV)
luminosity of RXJ1236.2+2558 is $\approx 3.5\times10^{39}$ erg s$^{-1}$. 
It coincides with $3.2\times10^{39}$ erg s$^{-1}$ from the
ROSAT PSPC observation (Vogler, Pietsch, \& Kahabka 1996). The absorption-corrected
X-ray luminosity in the 0.5-10 keV
band is $\approx 6.5\times10^{39}$ erg s$^{-1}$, which is
about one third of the value $2.3\times10^{40}$ erg s$^{-1}$ obtained
using the ASCA data (Mizuno et al. 1999). In comparison, the luminosity
for the central source in the ASCA data is about twice of
that in the Chandra data. So there is some evidence that both the
nucleus and the off-center source are variable,
although the variability needs to be established more carefully as
the spatial resolution of ASCA is insufficient to resolve these two sources.

Taking into account the extinction correction of 
A$_B$ as 0.6 (see section 3.3.2), the B magnitude of RXJ1236.2+2558 
is 24.5, which implies an X-ray to optical flux ratio (f$_{X}$/f$_{B}$)
of about 540. This value is similar to
that of the brightest ULX in M82 (Stocke, Wurtz \& K\"uhr 1991).

\subsubsection{Is RXJ1236.2+2558 a Foreground or Background Object?}

If we assume the optical counterpart of RXJ1236.2+2558 is a foreground
star in Milky Way, it would be an F type star according to its color. Then, its 
optical luminosity implies a distance of 100 kpc, which is at the edge of the halo 
of our Galaxy. More importantly, the high X-ray to optical flux ratio of 
RXJ1236.2+2558 is outside the observed range ($10^{-4} -
10^{-1}$) for normal B to M type stars where the X-ray emission
arises from a hot corona (Maccacaro al. 1988). The value
is also outside the range (0.1-10)  
for cataclysmic variables (Bradt \& McClintock 1983). Only Low 
Mass X-ray Binary systems (LMXBs) could reach such a high ratio. From
Bradt \& McClintock (1983), most LMXBs with 
high X-ray to optical flux ratios 
are distributed close
to the Galactic disk. The high Galactic latitude location 
(86.44$\rm ^{\circ}$) of the source 
disfavors this association. Furthermore, as Mizuno et al. (1999) pointed out, the
ASCA spectrum of RXJ1236.2+2558 is softer than those of LMXBs and hence
the possibility that RXJ1236.2+2558 is a LMXB can be ruled out.
Therefore, the optical counterpart of RXJ1236.2+2558 is unlikely to be
an object in the Milky Way.

On the other hand, if RXJ1236.2+2558 is a background extra-galactic
source, then the object can only be
a BL Lac object from its very high X-ray to 
optical flux ratio. However, the ratio for BL Lac objects is still
about one order of magnitude lower than that of the off-center source 
(Maccacaro et al. 1988). 
We emphasize that the observed $f_{\rm X}/f_{\rm B}$ cannot be made
compatible with those of BL Lacs through extinction, as it requires
$A_B \approx 3$ mag. Such a high extinction is inconsistent with
$A_B \approx 0.6$ mag estimated using $N_{\rm H}/E(B-V)=5.8
\times 10^{21} {\rm cm}^{-2}$  (Bohlin, Savage \& Drake 1978)  
and $A_B=4E(B-V)$ for $N_{\rm H} \approx 0.8 \times$10$^{21}$ 
cm$^{-2}$ (appropriate for the MCD model, Table 2). In addition, no radio 
emission has been detected at the position of RXJ1236.2+2558 by VLA (Ho \&
Ulvested 2001). Therefore, RXJ1236.2+2558 is unlikely to be a background BL Lac
object. Combining all the
evidence, we conclude that RXJ1236.2+2558 is probably not
a background extra-galactic object.

\subsubsection{A ULX in NGC~4565}

Given that RXJ1236.2+2558 is neither a foreground star nor a background
AGN, we are left with only one possibility, i.e., this source
is an X-ray luminous point source in NGC~4565. The X-ray luminosity 
in the 0.5-10 keV band is 6.5$\times$10$^{39}$ \erg, placing it
in the category of ULXs in a nearby normal spiral galaxy.

At a distance of 10 Mpc, the faint optical counterpart of RXJ1236.2+2558
has $M_B=-4.9$ and a color index of $B-I \approx 1.1 \pm 0.5$; both numbers
have not been corrected for reddening. Candidates for
point-like objects with $B\approx -5.0$ are only O stars (with type
later than O7) or globular
clusters. However, the optical counterpart of RXJ1236.2+2558 is very
unlikely to be an O star as such stars have much bluer colors
($B-I \approx -0.8$ instead of $B-I \approx 1.1$, as observed); the conclusion
is unchanged even if we allow reddening ($A_B \approx 0.6$). Moreover,
we do not expect an O star in the outer region of the bulge.
Similarly, the object is unlikely to be a very compact
star formation region, as these are usually found close to galactic disks.
Another possibility is that the optical counterpart is 
associated with a dwarf galaxy.  Its color is consistent with
that of some dwarf galaxies in the local group, however its luminosity and
size are both too small compared with those of dwarf galaxies
(e.g., Table 4 in Mateo 1998). One speculative scenario
may be that the optical counterpart
is the core of a tidally-stripped dwarf galaxy with
the AGN activity triggered by the tidal interaction.

Overall, we favor the interpretation that
the optical counterpart of ULX RXJ1236.2+2558 is probably a 
globular cluster.
It is interesting to compare the color and absolute
magnitude of this counterpart with those of globular clusters.
Most globular clusters 
have colors redder than $B-I=1.4$ (see Fig.~3 in Kissler-Patig et al. 1999).
The color of the object is $B-I=1.1$,
which is quite blue. However, with the large error-bar ($\approx 0.5$
mag) due to its faintness, its color is still consistent 
with being at the blue tail of the color distribution for globular clusters. The B-band
absolute magnitude of globular clusters is known to follow roughly a 
(universal) Gaussian distribution with a mean of $M_B=-7.1$ and a standard deviation of
1.1 mag. The extinction-corrected absolute magnitude with 
$M_B=-5.5\pm 0.4$ of the optical counterpart for RXJ1236.2+2558 is 
consistent with being at the faint end of the luminosity function
 (see Fig. 5 in Della Valle et al. 1998 and references therein).
 We conclude that the off-center X-ray source is hosted by a
globular cluster. We discuss the implication of this association
in more detail in the next section.

\section{Summary and Discussion}

In this paper, we have shown that, from multi-waveband observations in
the optical, radio and X-ray, the central X-ray source in NGC~4565 is
entirely consistent with it
being a low-luminosity AGN. We also argued that the off-center X-ray
point source is a ULX and its optical counterpart is  
a globular cluster as inferred from its color and absolute magnitude. 

Typical globular clusters in our galaxy have X-ray luminosities ranging up to 
~ 5\,$\times$10$^{37}$erg s$^{-1}$ (Ghosh et al. 2001). There are no
globular clusters with X-ray luminosities above $10^{38}$ erg s$^{-1}$ in M~31 (Supper et al. 
1997). NGC~4565 is an Sb galaxy similar to our galaxy, but it seems to
host a globular cluster that is about two orders of magnitude more luminous in
the X-ray than those found in the Galaxy and M31.
It is not clear what determines the upper cutoff in the X-ray
luminosity function of globular clusters in disk galaxies. In this regard, the upper cutoff
in NGC~4565 is similar to those seen in two
elliptical galaxies NGC~4697 and NGC~1399 (Sarazin, Irwin \& Bregman 
2001; Angelini, Loewenstein \& Mushotzky 2001). For example,
Angelini et al. found two of the three brightest point sources in
NGC~1399 (with $L_x \approx 5 \times 10^{39} {\rm erg~s^{-1}}$)
are hosted by globular clusters. At present, it is not
known whether the extreme X-ray luminosity is a sum of many LMXBs or
it is produced by an intermediate mass black hole. Angelini et
al. argued that the spectra of the two most luminous X-ray
globular clusters are consistent with those seen in
either high-state or low-state black holes in our Galaxy.

The formation and evolution of ULXs is still an unsolved problem. Models 
proposed so far can be classified into two categories, i.e. unbeamed models and beamed 
models. The beamed models suggest that ULXs represent a short-lived phase
in their evolution and are associated with young stellar populations, as
observed for many ULXs (see the introduction). However, for
RXJ1236.2+2558, this model does not work, as it is apparently in a
globular cluster. The unbeamed models, on the other 
hand, require an intermediate-mass BH ($\geq 100M_{\odot}$) in a binary with
an evolved donor star (King et al. 2001). However, the formation of such 
systems is uncertain. An intermediate- and high-mass BHs could be 
formed either through the merger of lower mass 
BHs in dense clusters or from early generation of zero-metallicity stars 
(so-called population III stars, Madau and Rees 2001). It is also possible that 
$\ge50 M_{\odot}$ population III BHs could form during the formation of
globular clusters; they will then sink 
to the center of the cluster quickly and grow to $\sim 10^3 M_{\odot}$
by accretion in  the cluster's lifetime (Miller \& Hamilton 2001). However, 
moderate-mass BHs may be difficult to retain in off-nuclear star 
clusters such as RXJ1236.2+2558 due to
the shallow potential well of globular clusters (Miller \& Hamilton 2001
and references therein).

No matter what is the origin of ULXs, it is important to further study them
in multi-wavebands. Extreme objects like RXJ1236.2+2558 offer
ideal targets for such studies.
As the exposure times for Chandra (2~ks) and HST ($\sim 500$s) are both
quite short for RXJ1236.2+2558, it is
important to obtain deeper images in both wavebands. 
Optical observations with HST (particularly with the Advanced Camera for
Surveys) will be useful to establish the color and the spatial
extent of the optical counterpart. On the other hand, deeper
integrations of Chandra will be useful to firmly establish the
spectral behavior of the off-center ULX, and its variability. The X-ray
observations may also be used to probe the X-ray corona seen in 
some edge-on galaxies (e.g., NGC4631, Wang et al. 2001). These
observations are important for further understanding the nature of ULXs, 
and the puzzle why some disk galaxies have ULXs while others 
(like the Milky Way and M31) have none.

\acknowledgments

We thank Drs. Luis Ho,  Simon White, Richard James and Renxin Xu for valuable 
discussion. Many thanks are also due to Drs. Xu Zhou, Jianyan Wei, Jifeng Liu, 
Jun Ma and Zhaoji Jiang. This work is supported by the Chinese 
National Natural Science Foundation (CNNSF) and NKBRSF G19990754.
S. M. acknowledges the financial support for his travel by the CNNSF.

\newpage
\begin{table}[ht]
\tablenum{1}
\caption[]{Positions (in J2000) for the central and off-center
 X-ray point sources from X-ray (Chandra), optical (HST) and radio (VLA)
at 6cm and 20cm}
\vspace {0.5cm}
\begin{tabular}{lcccc}
\hline
\hline
 &\multicolumn{2}{c}{RXJ1236.3+2559}  & \multicolumn{2}{c}{RXJ1236.2+2558} \\
 &RA & Dec & RA &Dec \\
\hline
\\
X-ray   &12:36:20.779 &25:59:15.74 &12:36:17.402 &25:58:55.44\\
\\
Optical &12:36:20.785  &25:59:16.04 &12:36:17.404 &25:58:55.80  \\
\\
Radio &12:36:20.781 &25:59:15.64 & ... & ...\\
\\
\hline
\end{tabular}\\
\end{table}

\newpage
\begin{table}[th]
\tablenum{2}
\caption[]{Best-fit spectral parameters for the X-ray spectra of
the central and off-center sources in NGC~4565. Two models are presented, an absorbed
multi-color disk model (MCD) and an absorbed power-law model.
The absorption column density (third column) is in units of $10^{21}~{\rm cm}^{-2}$.
The fourth column indicates either the inner-disk temperature ($T_{\rm
in}$) for the MCD model or the photon index for the power-law model.
The absorption-corrected fluxes in the 0.1-2.4 keV and 0.5-10 keV bands 
are both in units of $10^{-13}$ \fxu. 
The error bars in the Table are 90\% confidence limits.}
\vspace {0.5cm}
\begin{tabular}{ccccccc}
\hline
\hline
Source & Model & Absorption  & T$_{\rm in}~ {\rm or}~ \Gamma $ &
f(0.1-2.4 keV)& f(0.5-10 keV) & $\chi^2/{\rm dof}$\\
\hline
\\
Center     & MCD & $<1.6$ & 1.2$^{+0.8}_{-0.4}$ & 1.2$^{+2.4}_{-0.2}$  & 2.1$^{+4.1}_{-0.5}$  & 15.4/18 \\
\\
           & power-law & 1.1$^{+1.0}_{-0.9}$ & 1.6$^{+0.4}_{-0.4}$ & 2.1$^{+0.4}_{-0.4}$  & 3.6$^{+0.6}_{-0.6}$& 16.3/18  \\
\\
Off-center & MCD & $<0.8$ & 1.1$^{+0.4}_{-0.2}$ & 2.9$^{+4.3}_{-1.8}$  & 5.4$^{+7.3}_{-3.4}$ & 21.8/22 \\
\\
           & power-law & 1.5$^{+0.5}_{-0.3}$ & 1.8$^{+0.2}_{-0.2}$ & 7.1$^{+0.7}_{-0.7}$  & 7.5$^{+0.8}_{-0.8}$ & 20.2/22 \\
\hline
\end{tabular}
\end{table}

\newpage
\begin{table}[ht]
\tablenum{3}
\caption[]{Data for the nucleus of NGC~4565}
\vspace {0.5cm}
\begin{tabular}{cccc}
\hline
\hline
 $\nu$ & $\nu$$\rm f_{\nu}$& Aperture & Reference \\
 (Hz) & erg s$^{-1}$ cm$^{-2}$ & ($''$) &  \\
\hline
\\
 1.42 $10^9$& 3.27 $10^{-17}$ & 0.5$''$ & 1 \\
\\
 4.86 $10^9$& 1.31 $10^{-16}$ & 0.5$''$ & 1 \\
\\
 5.0 $10^9$& 1.25 $10^{-16}$ & 0.5$''$ & 2 \\
\\
 8.4 $10^9$& 2.10 $10^{-16}$ & 0.5$''$ & 2 \\
\\
 1.5 $10^{10}$& 4.65 $10^{-16}$ & 0.5$''$ & 2 \\
\\
 2.97 $10^{13}$& $<$ 2.55 $10^{-11}$ & 6.0$''$ & 3 \\
\\
 1.36 $10^{14}$& 8.43 $10^{-11}$ & 5.0$''$ & 4 \\
\\
 1.82 $10^{14}$& 1.93 $10^{-10}$ & 5.0$''$ & 4 \\
\\
 2.40 $10^{14}$& 1.18 $10^{-10}$ & 5.0$''$ & 4 \\
\\
 3.75 $10^{14}$& 2.50 $10^{-12}$ & 0.5$''$ & 5 \\
\\
 6.58 $10^{14}$& 6.93 $10^{-13}$ & 0.5$''$ & 5 \\
\\
 1.30 $10^{15}$& $<$ 3.00 $10^{-12}$ & 22$''$ & 6 \\
\\
 2.42 $10^{17}$&  8.1 $10^{-14}$ & 2.5$''$ & 7 \\
\\
\hline
\end{tabular}\\
(1) Ho \& Ulvestad 2001. (2) Nagar, Wilson \& Falcke 2001. 
(3) Rieke \& Lebofsky 1978. (4) measured from J,H,K images (Rice et al. 1996).
(5) measured from HST archive images (Kissler-Patig et al. 1999) 
(6) Maoz et al. 1996 (7) This paper
\end{table}

\clearpage
\begin{figure}
\figurenum{1}
\epsscale{.60}
\plotone{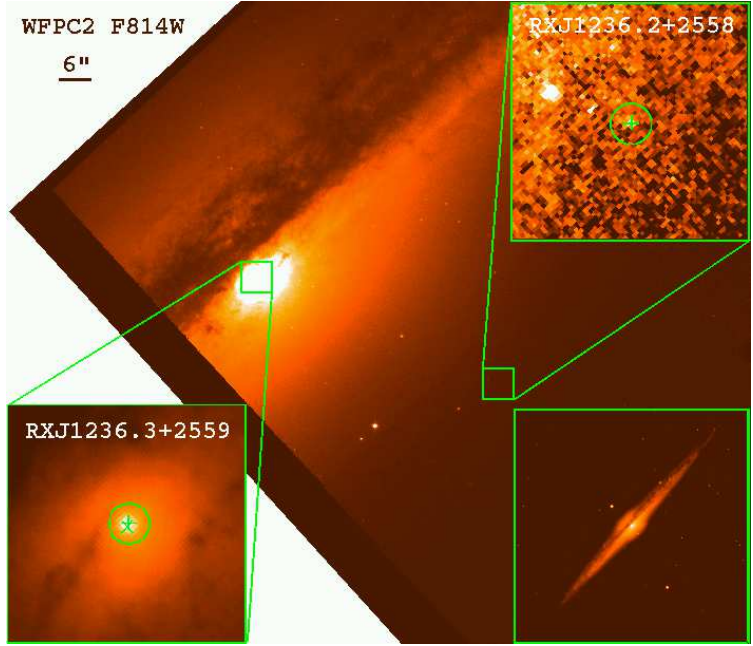}
\figcaption[fig1]{
HST WFPC2 image of the bulge of NGC~4565 in the F814W filter.  
The positions of X-ray sources RXJ1236.3+2559 and 
RXJ1236.2+2558 are indicated by two boxes. The bottom-left and top-right insets present zoomed
views of the nuclear X-ray source RXJ1236.3+2559 and the off-center X-ray source
RXJ1236.2+2558; both are shown as a plus with an error circle of 0.5\arcsec.
The cross symbol is the 6cm and 20cm radio position from Ho \& Ulvestad
(2001).
The coordinate system of the WFPC2 F814W image was shifted to match that of Chandra.
 From the zoomed images, we find that both X-ray sources have optical counterparts and 
the position agreement between optical and X-ray sources is better than
0.1\arcsec, within the absolute astrometric error of Chandra and HST.
The bottom right inset shows the whole image of NGC~4565 from the 60/90cm 
Schmidt telescope of NAOC.
\label{fig1}}
\end{figure}

\clearpage
\begin{figure}[t]
\figurenum{2}
\epsscale{.60}
\plotone{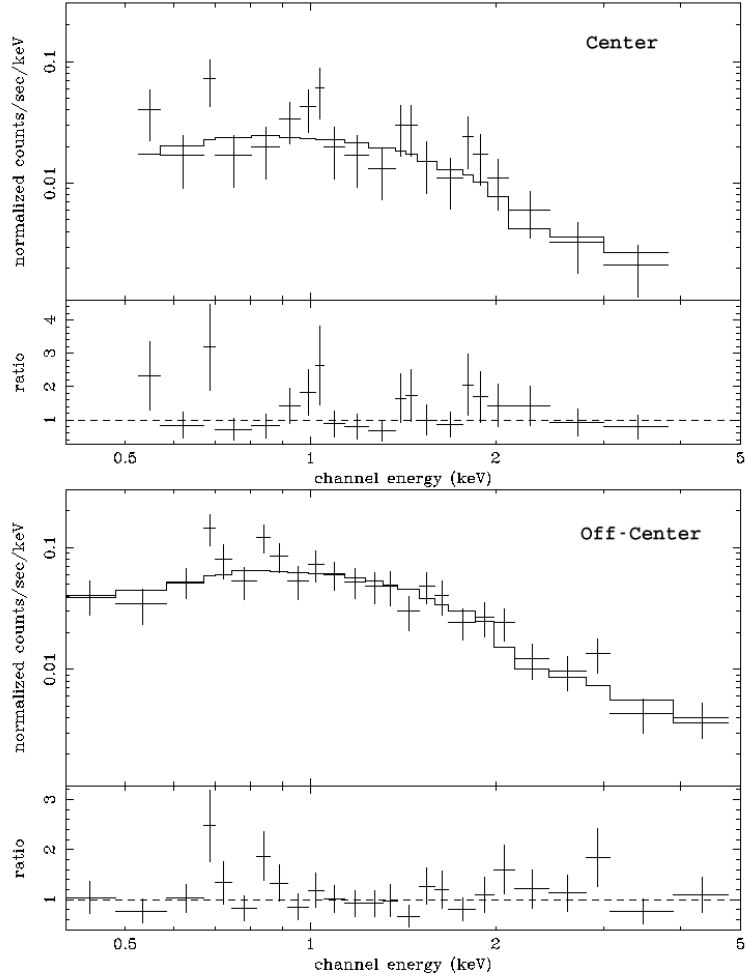}
\figcaption[fig2]{X-ray spectra for the 
nuclear source RXJ 1236.3+2559 (upper panel)
and the off-center source RXJ 1236.2+2558 (lower panel) in NGC~4565.
The spectrum of the nuclear source is fitted by 
an absorbed power-law and the off-nucleus source is fitted by an 
absorbed multi-color disk model. The fitting parameters are listed in Table 2.
\label{fig2}}
\end{figure}

\clearpage
\begin{figure}
\figurenum{3}
\plotone{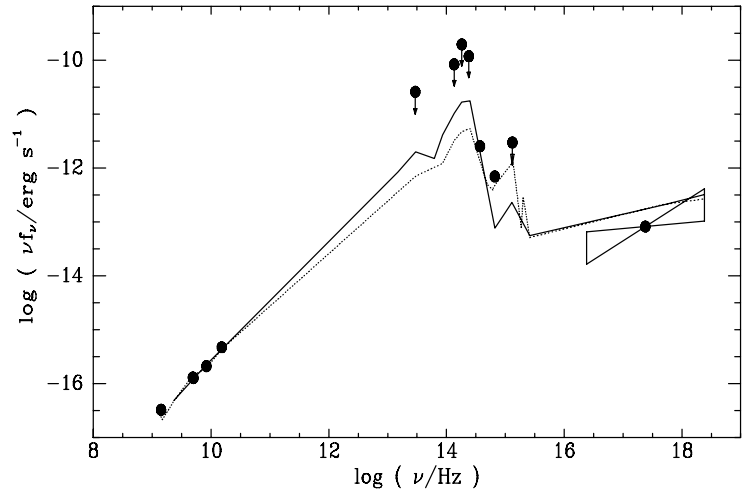}
\figcaption[fig3]{Spectral energy distribution for the
nucleus of NGC~4565 (see Table 3).
The solid and dotted lines are the spectral energy distributions
for the nuclei of NGC~4579 and M81 from Ho (1999), respectively.
For the purpose of comparison, the curves have been shifted downward by
$-1.0$ and $-1.5$, respectively. Upper limits are indicated by points
with arrows (see the text in \S3.3.3).
\label{fig3}}
\end{figure}

\end{document}